\documentclass[twocolumn,trackchanges]{aastex701}

\usepackage{amsmath}

\begin{document}

\title{Hierarchical cosmological constraints through strong lensing distance ratio}

\author[orcid=0000-0002-5594-0935]{Shuaibo Geng}
\affiliation{National Centre for Nuclear Research, Pasteura 7, PL-02-093 Warsaw, Poland}
\email{Shuaibo.Geng@ncbj.gov.pl}  

\author[orcid=0000-0002-8870-981X]{Shuo Cao$^{\ast}$}
\affiliation{School of Physics and Astronomy, Beijing Normal University, Beijing 100875, China}
\affiliation{Institute for Frontiers in Astronomy and Astrophysics, Beijing Normal University, Beijing 102206, China}
\email[show]{caoshuo@bnu.edu.cn} 

\author[orcid=0000-0003-1308-7304]{Marek Biesiada$^{\dagger}$}
\affiliation{National Centre for Nuclear Research, Pasteura 7, PL-02-093 Warsaw, Poland}
\email[show]{Marek.Biesiada@ncbj.gov.pl} 

\author{Xinyue Jiang}
\affiliation{School of Physics and Astronomy, Beijing Normal University, Beijing 100875, China}
\affiliation{Institute for Frontiers in Astronomy and Astrophysics, Beijing Normal University, Beijing 102206, China}
\affiliation{National Centre for Nuclear Research, Pasteura 7, PL-02-093 Warsaw, Poland}
\email{jiang.xinyue@foxmail.com} 

\author{Yalong Nan}
\affiliation{School of Physics and Astronomy, Beijing Normal University, Beijing 100875, China}
\affiliation{Institute for Frontiers in Astronomy and Astrophysics, Beijing Normal University, Beijing 102206, China}
\email{nanyalong2022@163.com} 

\author{Chenfa Zheng}
\affiliation{School of Physics and Astronomy, Beijing Normal University, Beijing 100875, China}
\affiliation{Institute for Frontiers in Astronomy and Astrophysics, Beijing Normal University, Beijing 102206, China}
\email{zcf@mail.bnu.edu.cn}

\begin{abstract}

Strong gravitational lensing provides an independent and powerful probe of cosmic expansion by directly linking observables to cosmological distances. Upcoming surveys such as LSST will discover large number of galaxy-galaxy strong lensing systems, offering a new route to precise cosmological constraints. In this paper, we propose a Fisher-like sensitivity factor to map how the cosmological information of strong-lensing distances changes across the lens-source redshift plane. Applying such factor to the distance ratio $D_{ls}/D_s$, the time-delay distance $D_{\Delta t}$, and the double-source-plane ratio, we determine the ``sensitivity valleys'' where an observable becomes insensitive to a given parameter. The realistically simulated LSST lens population, which largely lies outside the distance-ratio valleys, covers the most sensitive region for $(w_0,w_a)$ parameter space. We then develop a new hierarchical framework, which could calibrate the redshift evolution of lens mass-density slopes and constrain cosmological parameters simultaneously. Focusing on the LSST mock data, we demonstrate that ignoring mass-profile evolution can bias $\Omega_m$ by up to $\sim 10\sigma$, while modeling the lens evolution could perfectly recovers the fiducial cosmology and yield stringent cosmological constraints (e.g., $\Delta\Omega_m \simeq 0.01$ and $\Delta w \simeq 0.1$ for $\sim 10^4$ lenses).

\end{abstract}



\section{Introduction} 
\label{sec:intro}
The prevailing “vanilla model,” $\Lambda$ cold dark matter ($\Lambda$CDM), traces cosmic history from an early inflationary epoch, through the formation of galaxies to the present era of accelerated expansion. It accounts for a wide range of observations, most notably the precision measurements of the cosmic microwave background (CMB) \citep{planckcollaboration2020planck}. However, in recent years we have seen hints that this picture might be incomplete. One example is the Hubble tension \citep{freedman2021measurements, divalentino2021realm}, the persistent discrepancy between early universe inferences and late-universe distance-ladder measurements of $H_0$. In addition, analyses based on Dark Energy Spectroscopic Instrument (DESI) data release 2 (DR2) have been interpreted as showing a mild preference for a time-evolving dark-energy equation of state \citep{desicollaboration2025desi}. Hints like these motivate the pursuit of independent methods having distinct systematics to probe the expansion history of the Universe. One of such alternative and complementary method is based on strong gravitational lensing (SGL). As a particularly valuable cosmological probe, SGL can provide independent constraints on cosmological parameters. Rooted in the theory of general relativity (GR), SGL describes the deflection of light from a distant source by the gravitational field of a massive intervening object. This phenomenon offers a unique information about the link between the structure of galaxies and cosmological information, making it a powerful tool for probing dark matter \citep{2021MNRAS.502L..16C,2022A&A...659L...5C} and general relativity \citep{2017ApJ...835...92C,2020ApJ...888L..25C,2022ApJ...941...16L}.

There are three standard distance combinations for cosmological tests with strong lenses: the distance ratio $D^A_{ls}/D^A_s$ \citep{biesiada2010cosmic,cao2012constraintsa,cao2015cosmology,wang2020cosmological,qi2022cosmological,li2024cosmology}, where $D^A_{ls}$ is the angular diameter distance between lens and the background source, while $D^A_s$ is the angular diameter distance between source and the observer. The time-delay distance $D_{\Delta t}$ \citep{wong2020h0licowa,birrer2020tdcosmo,tdcosmocollaboration2025tdcosmo}, and the double–source–plane (DSP) ratio   \citep{collett2012constraining,collett2014cosmological,sahu2025cosmography,bowden2025constraining} which is essentially the ratio of $D^A_{ls}/D^A_s$ ratios corresponding to different sources. These three combinations are inferred from image or time-domain observables and lens models, with no additional astrophysical calibrators. So they can constrain cosmological parameters in any specified cosmological model. With $\mathcal{O}(10^2)$ lenses, the distance ratio alone yields $\Omega_m = 0.381^{+0.185}_{-0.154}$ \citep{chen2019assessing}. For time-delay cosmography, a single well-modeled system reaches about $5\%\text{--}8\%$ precision on $D_{\Delta t}$. An ensemble of eight lenses gives $H_0 \simeq 71.6^{+3.9}_{-3.3}\,\mathrm{km\,s^{-1}\,Mpc^{-1}}$ under flat $\Lambda$CDM, which is a measurement of $\sim 5\%$ precision \citep{tdcosmocollaboration2025tdcosmo}. For the DSP, just one system can deliver $w = -1.52^{+0.49}_{-0.33}$ and $\Omega_m = 0.192^{+0.305}_{-0.131}$; combining three systems improves these constraints by about $15\%$ \citep{bowden2025constraining}.

For cosmological applications of the time-delay distance and DSP lenses, the time-delay distance is particularly sensitive to the Hubble constant $H_0$ and can benefit greatly from $\sim 1\%$ time-delay measurements of lensed transients. At present, the most up-to-date TDCOSMO sample includes eight time-delay lensed quasars \citep{tdcosmocollaboration2025tdcosmo}. In DSP systems, the mass-sheet degeneracy (MSD) can be partially broken using imaging data alone if the lensing contribution from the first source plane is negligible. In practice, however, stellar velocity-dispersion measurements are still needed to constrain the lens-galaxy density slope in both cases. For DSP lenses, they are especially important to further break the MSD when the first-source mass cannot be ignored. Until very recently, the ``Jackpot'' lens J0946+1006 \citep{Jackpot2008} was the only such system known, and it has since been confirmed as a triple-source-plane system. To date, about 10 DSP systems have been spectroscopically confirmed \citep{DSP2,DSP3,DSP4,DSP5,barone2025agel,bowden2025constraining,sahu2025cosmography}, and ongoing wide-area surveys offer strong prospects for many more discoveries. The $D^A_{ls}/D^A_s$ distance ratio, as a $H_0$-independent distance combination, has a degeneracy direction in parameter space that notably differs from that of the CMB, so combining it with CMB constraints can break parameter degeneracies and sharpen inferences \citep{piorkowska2013complementarity}. Another advantage of galaxy-scale distance ratio is the sample size. Compared with lens systems that yield reliable time-delay distances and with DSP systems, galaxy–galaxy lenses with early-type deflectors form a much larger population. Velocity-dispersion measurements, however, add a significant uncertainty, which propagates into the cosmological constraints. Even so, if the galaxy-scale lens sample is properly calibrated at the population level, distance-ratio cosmography remains a promising and competitive probe.

Various previous studies reported a noticeable redshift evolution of the total mass–density slope of lens galaxies \citep{li2018stronglensing,etherington2022bulgehalo,tan2024joint}. Simulations also showed possible evolution of such kind  \citep{remus2017coevolution,peirani2019total}, but the direction of the evolution was not yet settled. Analyses that include stellar dynamics tended to prefer steeper slopes toward lower redshift \citep{bolton2012boss,sonnenfeld2013sl2sb,li2018stronglensing,chen2019assessing}. Image-only inferences often find flatter profiles from high redshift to more current epochs. \citep{etherington2022bulgehalo,tan2024joint}. \citet{etherington2022bulgehalo} suggest that the difference can arise from how the slope is defined at different physical scales. \citet{sahu2024agel} emphasized modeling systematics and sample limitations and argued for little or no evolution. This possible evolution of lens mass profile matters not only for tests of galaxy formation, but also affects cosmological inference. As shown in \citet{chen2019assessing,li2024cosmology}, ignoring the redshift evolution of lens mass distribution could bias cosmological constraints. This indicates a strong degeneracy between cosmological parameters and the evolution of the lens mass-density profiles.

In this work we implement a hierarchical cosmological inference for a galaxy-scale lens sample. First, we calibrate the population-level redshift evolution of the total mass–density slope and the luminous-matter density slope, without assuming any specific dark-energy model. This first tier was to some extent performed in the earlier paper \citep{geng2025investigating} where we used the unanchored Type~Ia supernovae (SN Ia) to Artificial Neural Network (ANN) reconstruct the cosmological distances as a function of redshift as suggested purely by the data. Let us stress that reconstructed distance ratios were  what we needed, hence the unanchored SN Ia were sufficient, bypassing the uncertainties related to the value of the Hubble constant. In the second tier, to which this paper is devoted, we use the effective power-law mass-density models with evolving slopes to infer the $D^A_{ls}/D^A_s$ distance ratios from strong lensing observables. We then compare the distance ratios from observables with the theoretical predictions to constrain cosmological parameters, with marginalized nuisance parameters such as velocity anisotropy. This workflow is able to yield cosmological constraints while controlling their degeneracy with redshift evolution. The structure of the paper is the following. Section \ref{sec:theory} sets the theoretical framework of the distance ratio cosmology in models beyond the standard $\Lambda$CDM model. Section \ref{sec:data} outlines the data used, including the observed data, the simulation data, and the data from \textit{Planck} CMB data for the joint constraints. In Section \ref{sec:Methodology}, we detail our analysis on the sensitivity of different cosmological parameters through distance ratio and the hierarchical inference framework to constrain cosmological parameters. Our approach to sensitivity is similar to (but not the same as) the one adopted in Dark Energy Task Force report \citep{DETF} based on the Fisher matrix approach. Therefore from now on we will refer to our method as Fisher-like. Section \ref{sec:results} presents our results, followed by the conclusions in Section \ref{sec:summary}.

\section{Theory}\label{sec:theory}
\subsection{Cosmological models}
Invoking the cosmological principle, a spatially homogeneous and isotropic universe is described by the Friedmann--Lema\^{i}tre--Robertson--Walker (FLRW) metric,
\begin{equation}\label{eq:FLRW}
ds^2 = dt^2 - a^2(t)\left[\frac{dr^2}{1 - k r^2} + r^2 \left(d\theta^2 + \sin^2 \theta\, d\varphi^2\right)\right],
\end{equation}
where $(r,\theta,\varphi)$ are comoving coordinates and $a(t)=1/(1+z)$ is the scale factor. In the following, we adopt $k=0$, corresponding to a spatially flat geometry. By modeling the cosmic contents as perfect fluids and including a cosmological constant $\Lambda$, which behaves as a uniform energy component with negative pressure $p=-\rho$, the $\Lambda$CDM model gives
\begin{equation}
\begin{aligned}
H^2 &= \frac{8\pi G}{3}\,\rho \;+\; \frac{\Lambda}{3} \\
&= H_0^2 \left[\Omega_m (1+z)^3  + \Omega_\Lambda\right],
\end{aligned}
\end{equation}
where $H_0$ is the Hubble constant, $\Omega_m \equiv {\rho_m}/{\rho_{\mathrm{cr}}}$ and $\Omega_{\Lambda} \equiv {\rho_{\Lambda}}/{\rho_{\mathrm{cr}}}$ are the density parameter for matter component and dark energy component, respectively, defined by the critical density $\rho_{\mathrm{cr}} \equiv {3H_0^2}/(8\pi G)$. Here we neglect radiation and massive neutrinos due to their negligible contribution to the expansion of the Universe at late times.

To go beyond $\Lambda$CDM, a minimal extension is the flat constant-$w$ dark energy model (wCDM), in which the dark-energy equation-of-state parameter $w \equiv p/\rho$ is allowed to differ from $-1$ but is assumed constant in time. For a non-interacting component with constant $w$,
\begin{equation}
\rho_{\mathrm{DE}}(a) \propto a^{-3(1+w)},
\end{equation}
and the Hubble parameter becomes
\begin{equation}
H^2 = H_0^2 \left[\Omega_m (1+z)^3 + \Omega_{\mathrm{DE}} (1+z)^{3(1+w)}\right],
\end{equation}
where $\Omega_{\mathrm{DE}} = 1 - \Omega_m$ (again neglecting radiation). The special case $w=-1$ reduces to $\Lambda$CDM. Cosmic acceleration requires $w<-1/3$. Values $-1<w\lesssim -1/3$ are typically associated with \emph{quintessence}-like behavior \citep{brax1999quintessence}, while $w<-1$ corresponds to the \emph{phantom} dark energy \citep{caldwell2002phantom}. The flat $w_0w_a$CDM model extends flat $w$CDM by allowing the dark-energy equation of state to evolve with time, using the Chevallier–Polarski–Linder (CPL) form \citep{chevallier2001accelerating,linder2003exploring} of
\begin{equation}
w(a)=w_0+w_a(1-a).
\end{equation}
This is a flexible, phenomenological parametrization. It is not tied to a single microphysical model, but it can approximate many physically motivated scenarios since it is essentially a first order Taylor expansion of any $w(z) \equiv{w(a(t))}$ around $a = 1$.

\subsection{Strong Lensing }\label{sec:SGL theory}
Regarding the phenomenological model of the lens, we include both baryonic matter and dark matter, which together dominate the galaxy mass distribution. We adopt the extended power-law mass-density model of \citet{koopmans2006gravitational}. The luminous tracer and the total mass follow power-law radial profiles, denoted by $\rho_{\mathrm{lum}}(r)$ and $\rho_{\mathrm{tot}}(r)$. The anisotropy parameter $\beta$ describing the three-dimensional velocity-dispersion anisotropy of the luminous tracer is also included as \citep{2016MNRAS.461.2192C}
\begin{equation}\label{eq:mass distribution assumption}
\begin{aligned}
\rho_{\mathrm{lum}}(r) &= \rho_{\mathrm{lum},0}\, r^{-\delta},\\
\rho_{\mathrm{tot}}(r) &= \rho_{\mathrm{tot},0}\, r^{-\gamma},\\
\beta(r) &= 1-\frac{\langle \sigma_{v,\theta}^2\rangle}{\langle \sigma_{v,r}^2\rangle}.
\end{aligned}
\end{equation}
Here, $\delta$ and $\gamma$ are the logarithmic density slopes of the luminous (baryonic) component and the total matter (dark matter plus baryons), respectively. 
We model their redshift evolution with a linear relation:
$\gamma=\gamma_0+\gamma_s\,(z_l-z_{l,\mathrm{med}})$ and
$\delta=\delta_0+\delta_s\,(z_l-z_{l,\mathrm{med}})$,
where $z_{l,\mathrm{med}}$ is the median lens redshift of the sample. 
The anisotropy parameter $\beta$ is set by the ratio of the tangential velocity dispersion $\sigma_{v,\theta}$ to the radial velocity dispersion $\sigma_{v,r}$. Upon solving the spherical Jeans equation under these assumptions, one can derive the dynamical mass $M_{dyn}$ of the lensing galaxy \citep{koopmans2006gravitational}, which corresponds to the projected mass enclosed within $\theta_E$,
\begin{equation} \label{eq:dynamical mass}
M_{dyn} = \frac{\pi}{G} \sigma_{v}^2 D^A_l \theta_E \left( \frac{\theta_E}{\theta_{ap}} \right)^{2-\gamma} f(\gamma,\delta,\beta),
\end{equation}
where
\begin{equation} \label{eq:general f function}
\begin{aligned}
f(\gamma,\delta,\beta)&= \frac{1}{2\sqrt{\pi}}\left(\frac{(\gamma+\delta-5)(\gamma+\delta-2-2\beta)}{\delta-3} \right) \\
&\times \frac{\Gamma\left( \frac{\gamma+\delta-2}{2}\right)\Gamma\left( \frac{\gamma+\delta}{2}\right)}{\Gamma\left( \frac{\gamma+\delta}{2}\right)\Gamma\left( \frac{\gamma+\delta-3}{2}\right)-\beta \Gamma\left( \frac{\gamma+\delta-2}{2}\right)\Gamma\left( \frac{\gamma+\delta-1}{2}\right)} \\
&\times  \frac{\Gamma\left( \frac{\delta-1}{2}\right)\Gamma\left( \frac{\gamma-1}{2}\right)}{\Gamma\left( \frac{\delta}{2}\right)\Gamma\left( \frac{\gamma}{2}\right)},
\end{aligned}
\end{equation}
and $\Gamma$ denotes the well-known Gamma function. The formula above displays the mass inferred from the luminosity-weighted average line-of-sight velocity dispersion $\sigma_{v}$ within the specified aperture $\theta_{ap}$ of the spectrograph and illustrates the usefulness of the power-law profile to scale between $\theta_E$ and $\theta_{ap}$. In practice, constant $\theta_{ap}$ would correspond to different physical scales in different galaxies, hence, as it will be discussed in details later one adjusts the velocity dispersion to the half of the effective radius of given galaxy. The value of $\beta = 0$ indicates an isotropic, spherically symmetric density distribution, and setting both $\gamma$ and $\delta$ to 2 recovers the singular isothermal sphere (SIS) model. On the other hand, the mass inside the Einstein radius could be robustly recovered from the strong lensing images
\begin{equation} \label{lensing mass}
 M_{lens} = \frac{c^2}{4 G} \frac{D^A_l D^A_s}{D^A_{ls}} \theta_E^2
\end{equation}
Then we can define a distance ratio as
\begin{equation} \label{eq:lensing distance ratio}
\mathcal{D} \equiv \frac{D^A_{ls}}{D^A_s}=\frac{c^2}{4 \pi } \frac{\theta_E}{\sigma_{v}^2} \left( \frac{\theta_E}{\theta_{ap}} \right)^{\gamma - 2} f^{-1}(\gamma),
\end{equation}
with the absolute uncertainty
\begin{equation} \label{theo uncertainty}
\Delta\mathcal{D}_i=\mathcal{D}_i\sqrt{4(\delta\sigma_{v})^2+(1-\gamma)^2(\delta\theta_E)^2},
\end{equation}
obtained by propagation of uncertainties in measurements of velocity dispersion and the Einstein radius. Fractional uncertainties $\delta\sigma_{v}$ and $\delta\theta_E$ will be discussed in Section \ref{sec:data}. In our analysis, we disregard the covariance between velocity dispersion and Einstein radii, as these measurements are derived from different instruments and methodologies. For simplicity, we assume that there are no correlated uncertainties between them.


\section{Data} \label{sec:data}

\subsection{Current strong lensing sample}\label{subsec:SGL_sample}
In this work, we use a galaxy-scale SGLs sample, originally assembled by \citet{cao2015cosmology} and later updated by \citet{chen2019assessing}.\footnote{Historically, these compilations were preceded by earlier ones \citep{Biesiada2010,cao2012constraints} carefully selected from discovery papers. We avoided mechanistic use of then existing databases like CASTLES, even though this and other data bases are extremely useful.} This sample comprises 161 galaxies, predominantly early-type (E/S0 morphologies), carefully selected to exclude those with significant substructures or near companions. The dataset integrates 5 systems from the LSD survey \citep{koopmans2002stellar, koopmans2003structure, treu2002internal, treu2004massive}, 26 from the CFHTLS Strong Lensing Legacy Survey (SL2S)  \citep{ruff2011sl2s, sonnenfeld2013sl2sa, sonnenfeld2013sl2sb, sonnenfeld2015sl2s}, 57 from the Sloan Lens ACS Survey (SLACS) \citep{bolton2008sloan, auger2009sloan, auger2010sloanb}, 38 from the an extension of the SLACS survey known as “SLACS for the Masses (S4TM)” \citep{shu2015sloan, shu2017sloan}, 21 from the BELLS \citep{brownstein2012boss}, and 14 from the BELLS for GALaxy-Ly EmitteR sYstems (GALLERY) \citep{shu2016boss, shu2016bossa}. 

\begin{figure*}
\centering
\includegraphics[width=0.6\linewidth]{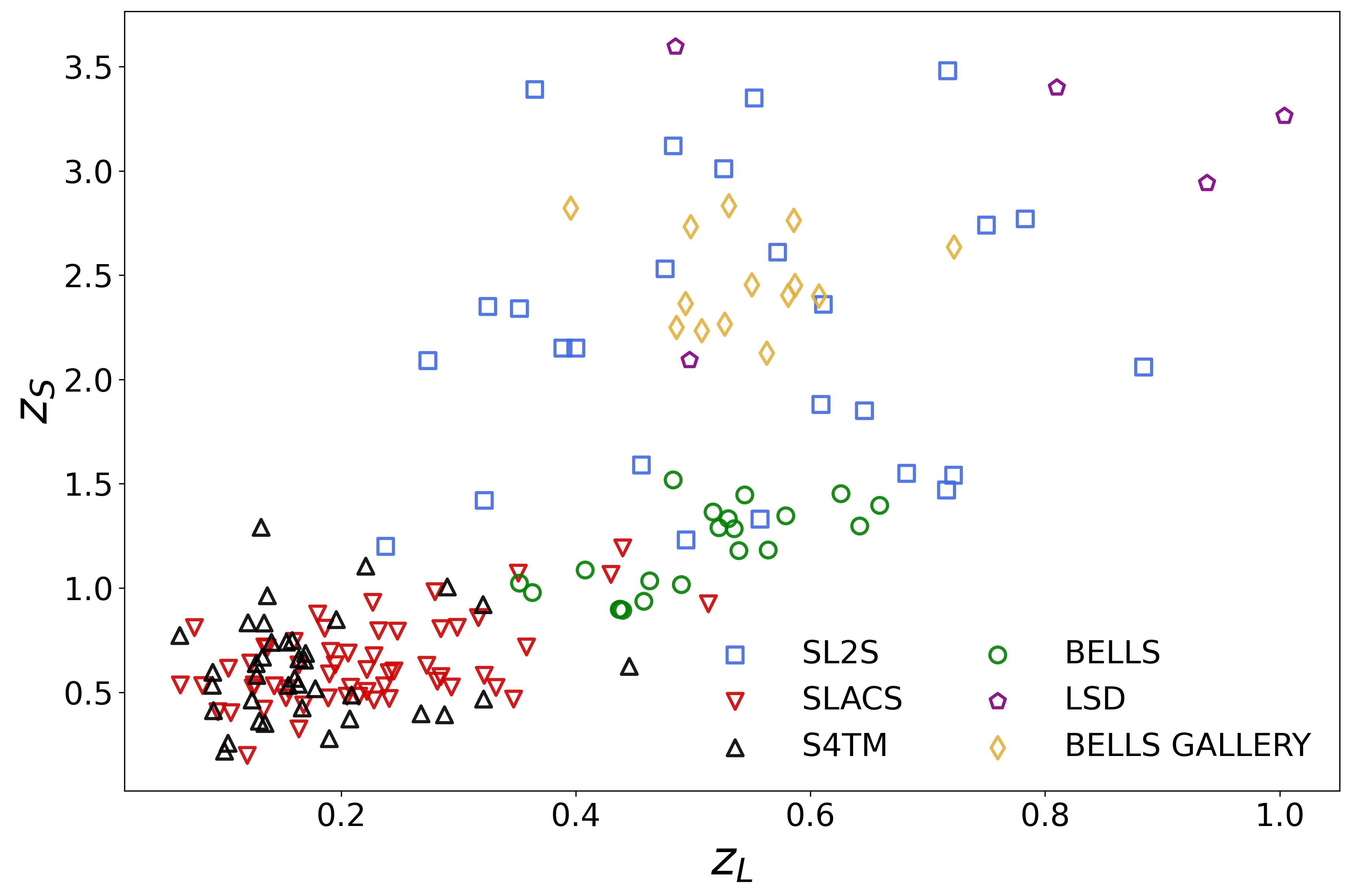}
\caption{\label{fig:scatter} Redshift distribution of lens galaxies and background sources in the combined sample used in this work. Blue squares represent lenses from SL2S, green circles represent lenses from BELLS, red down triangles represent lenses from SLACS, purple pentagons represent lenses from LSD, black triangles represent lenses from S4TM, and yellow diamonds represent lenses from BELLS GALLERY. }
\end{figure*}

For our analysis, we used the following observables: (1) spectroscopic redshifts of the lensing galaxies ($z_l$) and the sources ($z_s$); (2) the Einstein radius ($\theta_E$), with an assumed fractional uncertainty of $\delta\theta_E=5\%$, (3) the effective radius ($\theta_{\text{eff}}$), i.e. the half-light radius for lens galaxy; and (4) the velocity dispersion ($\sigma_{ap}$) within an aperture of angular radius ($\theta_{ap}$) along with its uncertainty. The redshift distribution of our combined sample are shown in Fig.~\ref{fig:scatter}. The velocity dispersion is measured using a rectangular slit, hence the equivalent circular aperture is calculated as \citep{jorgensen1995spectroscopy}
\begin{equation} \label{eq:effective aperture}
\theta_{ap}\simeq 1.025\times\sqrt{(\theta_x\theta_y/\pi)}.
\end{equation}
To ensure consistency across different physical sizes measured in the fixed-sized aperture, we normalized measured velocity dispersions to 
the apertures equivalent to half of the effective radius $\theta_{e/2}$, denoted by $\sigma_{v,e/2}$, employing the aperture correction formula from \citep{jorgensen1995spectroscopy}
\begin{equation} \label{eq:aperture correction}
\sigma_v^{\text{obs}} = \sigma_{e/ 2} = \sigma_{\text{ap}} \left( \frac{\theta_{\text{eff}}}{2\theta_{\text{ap}}} \right)^\eta.
\end{equation}
For the value of the correction parameter, we adopt $\eta=-0.066 \pm 0.035$, as determined by \citet{cappellari2006sauron}. Following \citet{chen2019assessing}, the total uncertainty in the velocity dispersion is calculated as
\begin{equation} \label{eq:VD uncertainty}
(\Delta \sigma_{e/2}^{\text{tot}})^2 = (\Delta \sigma_{e/2}^{\text{stat}})^2 + (\Delta \sigma_{e/2}^{\text{AC}})^2 + (\Delta \sigma_{e/2}^{\text{sys}})^2,
\end{equation}
where $\Delta \sigma_{e/2}^{\text{stat}}$ represents the statistical uncertainty from the measurements, $\Delta \sigma_{e/2}^{\text{AC}}$ is propagated uncertainty from the aperture correction to half effective radius, which is calculated as
\begin{equation} \label{eq:aperture correction uncertainty}
\begin{aligned}
(\Delta \sigma_{e/2}^{\text{AC}})^2 &= (\Delta \sigma_{ap})^2 \left(\frac{\theta_{eff}}{2 \theta_{ap}}\right)^{2 \eta} \\
&+  \sigma_{ap}^2 \left(\frac{\theta_{eff}}{2 \theta_{ap}}\right)^{2 \eta} \left( \ln{\frac{\theta_{eff}}{2 \theta_{ap}}}\right)^2 (\Delta \eta)^2,
\end{aligned}
\end{equation}
and $\Delta \sigma_{e/2}^{\text{sys}}$ accounts for the systematic uncertainty associated with the assumptions made regarding the consistency between $M_{dyn}$ and $M_{lens}$. We incorporate a systematic uncertainty of 3\% on the model-predicted velocity dispersion to account for the potential influence of line-of-sight contamination, as suggested by \citet{jiang2007baryon}. 
From the overall uncertainty mentioned in Eq.~(\ref{eq:VD uncertainty}), we can calculate the relative uncertainty $\delta\sigma_{ap}$ in Eq.~(\ref{theo uncertainty}) by $\delta\sigma_{ap} = \Delta \sigma_{e/2}^{\text{tot}} / \sigma_{e/2}$.

\subsection{Simulated strong lensing sample}\label{sec:simulation data}

In order to examine the possible degeneracy between cosmological parameters and the density slope evolutions, we generated mock datasets of galaxy-galaxy strong lensing observables based on Vera Rubin Observatory Legacy Survey of Space and Time (LSST) \citep{collett2015population,ivezic2019lsst} forecast models. To model the LSST lens population, we follow the methods of \citet{li2024cosmology,geng2025investigating} and the population forecast of \citet{collett2015population}. The fiducial cosmological parameters used in the simulation are listed in Table~\ref{tab:cosmo-input}. We simulate 7500 systems with spectroscopic information, consistent with spectroscopic galaxy--galaxy lens catalogs. From this SGL population, we take the lens redshifts, source redshifts, and Einstein radii, which we use to infer the velocity dispersion.
In our real-data SGL catalog described in \ref{subsec:SGL_sample}, we found an average relative error of 11\% for $\Delta \sigma_{e/2}^{\text{tot}}/ \sigma_{ap}$. Hence, we applied this relative uncertainty to the mock velocity dispersion values. During this process, we incorporated probability distributions for the slopes of the mass density ($\gamma$ and $\delta$) and the anisotropy parameter ($\beta$) based on the results presented in \citet{geng2025investigating}. 
Specifically, we adopted redshift evolution of the total mass density slope $\gamma = 2.054(\pm 0.042) - (z_l-z_{l,\mathrm{med}}) \times 0.19(\pm 0.11)$, and the stellar mass density slope $\delta =2.26(\pm 0.13) - (z_l-z_{l,\mathrm{med}}) \times 0.16(\pm 0.18)$, while setting $\beta \sim \mathrm{Tri}\big(-0.5,\,0.656;\ \mathrm{mode}=0.102\big)$.

In this analysis, we also use the \textit{Planck} 2018 cosmological results to assess joint constraints. We adopt the official $\mathrm{plikHM\_TTTEEE\_lowl\_lowE}$ and $\mathrm{plikHM\_TTTEEE\_lowl\_lowE\_BAO}$ MCMC chains from \citet{planckcollaboration2020planck}. The baryon acoustic oscillation (BAO) data included in $\mathrm{plikHM\_TTTEEE\_lowl\_lowE\_BAO}$ follow the default \textit{Planck} BAO compilation, combining measurements from the 6dF Galaxy Survey (6dFGS) \citep{beutler20116df}, the SDSS Main Galaxy Sample \citep{ross2015clustering}, and the Baryon Oscillation Spectroscopic Survey Data Release 12 (BOSS DR12)\citep{alam2017clusteringa}. We then label constraints based on this chain as ``Planck+BAO'' throughout. For dark energy models beyond flat $\Lambda$CDM (e.g., $w$CDM and $w_0w_a$CDM), we use the corresponding \textit{Planck} chains and combine them with the strong-lensing likelihood via importance reweighting, assuming statistical independence. This combination is effective because the CMB degeneracy directions differ from those of strong-lensing distance ratios.

\section{Methodology} \label{sec:Methodology}
This section describes our sensitivity analysis for different distance combinations and outlines the hierarchical framework used to jointly constrain the lens-population and cosmological parameters.

\subsection{Sensitivity analysis}\label{sec:sensitivity analysis}
Ongoing surveys such as LSST and Euclid \citep{euclidcollaboration2022euclid}, together with the upcoming China Space station Survey Telescope (CSST) \citep{csstcollaboration2025introduction,cao2024csst}, will discover large samples of galaxy–galaxy lenses, greatly improving distance-ratio constraints on cosmological parameters. With these larger samples, we will also identify more lensed-AGN systems for time-delay cosmology and double–source-plane systems, which are extremely rare in current datasets. 
To assess the sensitivity of these distance combinations to cosmological parameters, we will perform a Fisher-like analysis around the fiducial model
$\Omega_m=0.3,\ w_0=-0.95,\ w_a=0.0$, and define the sensitivity to the parameter $\lambda_i \in \{\Omega_m, w_0, w_a \}$ as
\begin{equation}
  \frac{1}{\sigma(\lambda_i)} \equiv
  \frac{\left|\frac{\partial \ln \mathcal{R}}{\partial \lambda_i}\right|}{\sigma_{\ln \mathcal{R}}},
\end{equation}
where $\lambda_i$ is the parameter of interest, and $\sigma(\lambda_i)$ -- its uncertainty.  $\mathcal{R}$ denotes the chosen distance combination: distance ratio ($\mathcal{D}$), time-delay distance ($\mathcal{D}_{\Delta t}$), or double-source-plane distance ratio ($\mathcal{D}_{DSP}$), respectively. The distance ratio is defined as in 
Eq.~\ref{eq:lensing distance ratio}.
The theoretical time-delay distance is
\begin{equation}
    \mathcal{D}_{\Delta t}= (1+z_l)\,\frac{D^A_l D^A_s}{D^A_{ls}}.
\end{equation}
While the theoretical double-source-plane distance ratio is 
\begin{equation}
\mathcal{D}_{DSP}=\frac{D^A_{ls_1}/D^A_{s_1}}{D^A_{ls_2}/D^A_{s_2}},
\end{equation}
where subscripts $s_1$ and $s_2$ denote the two sources.
Here, we ignore the external convergence and the internal mass-sheet degeneracy. We analyze the sensitivity of the different observables only from the point view of cosmology.

\begin{figure*}[ht]
\centering
\includegraphics[width=0.95\linewidth]{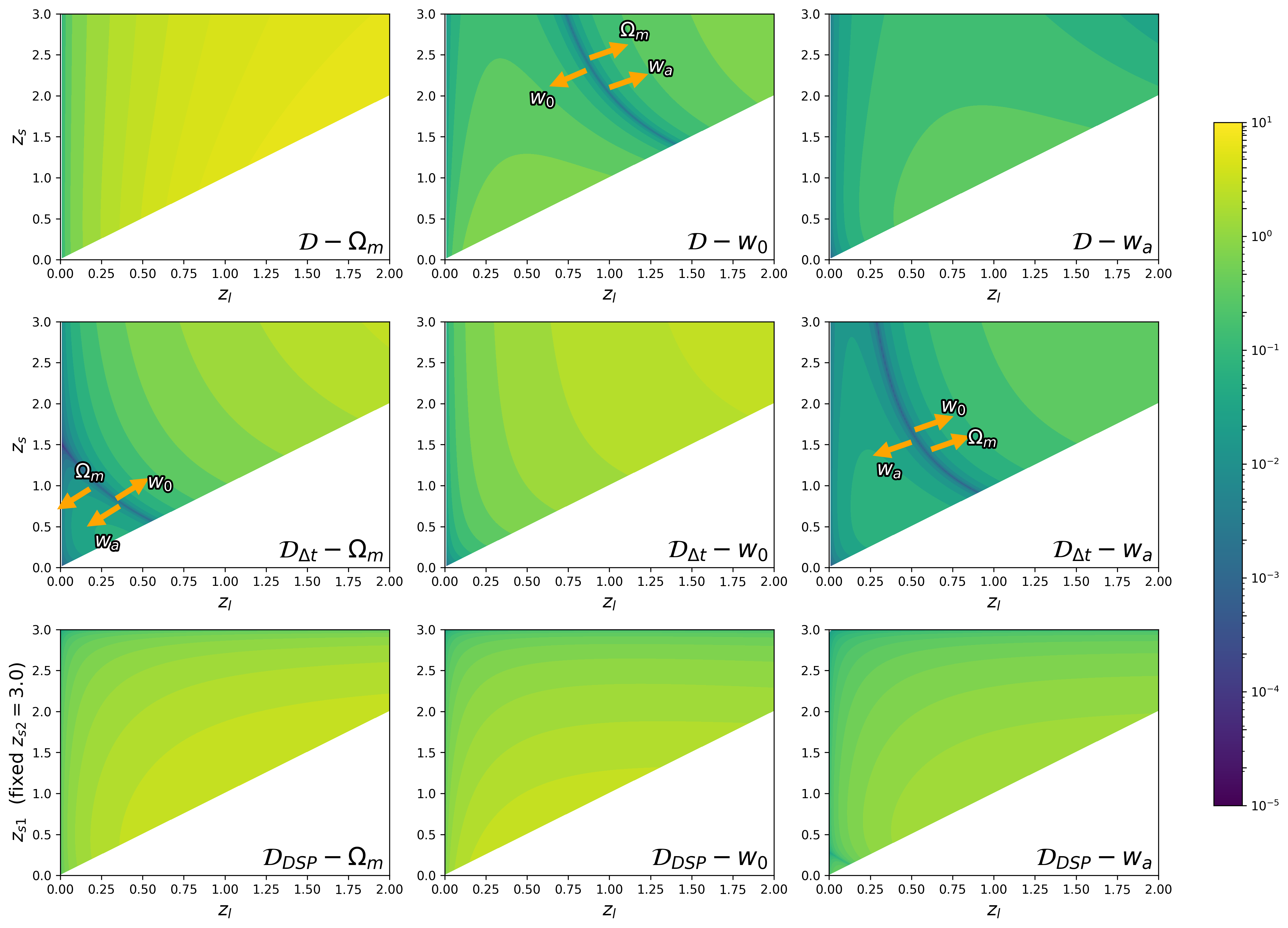}
\caption{\label{fig:fisher analysis}Sensitivity maps for the three cosmological parameters $\Omega_m$, $w_0$, and $w_a$. 
Top row: sensitivity factor of the distance ratio as a function of lens redshift $z_l$ and source redshift $z_s$. 
Middle row: sensitivity factor of the time-delay distance as a function of $z_l$ and $z_s$. 
Bottom row: sensitivity factor of the double–source–plane distance ratio with the higher-redshift source fixed at $z_{s2}=3.0$, shown as a function of $z_l$ and the lower-redshift source $z_{s1}$. 
Larger values correspond to stronger sensitivity of a given distance combination to the cosmological parameters, although the absolute normalization is not directly comparable between different distance combinations. 
The orange arrows indicate how the locations of the low-sensitivity valleys shift when a given parameter is increased.
}
\end{figure*}

\begin{figure*}
\centering
\includegraphics[width=0.95\linewidth]{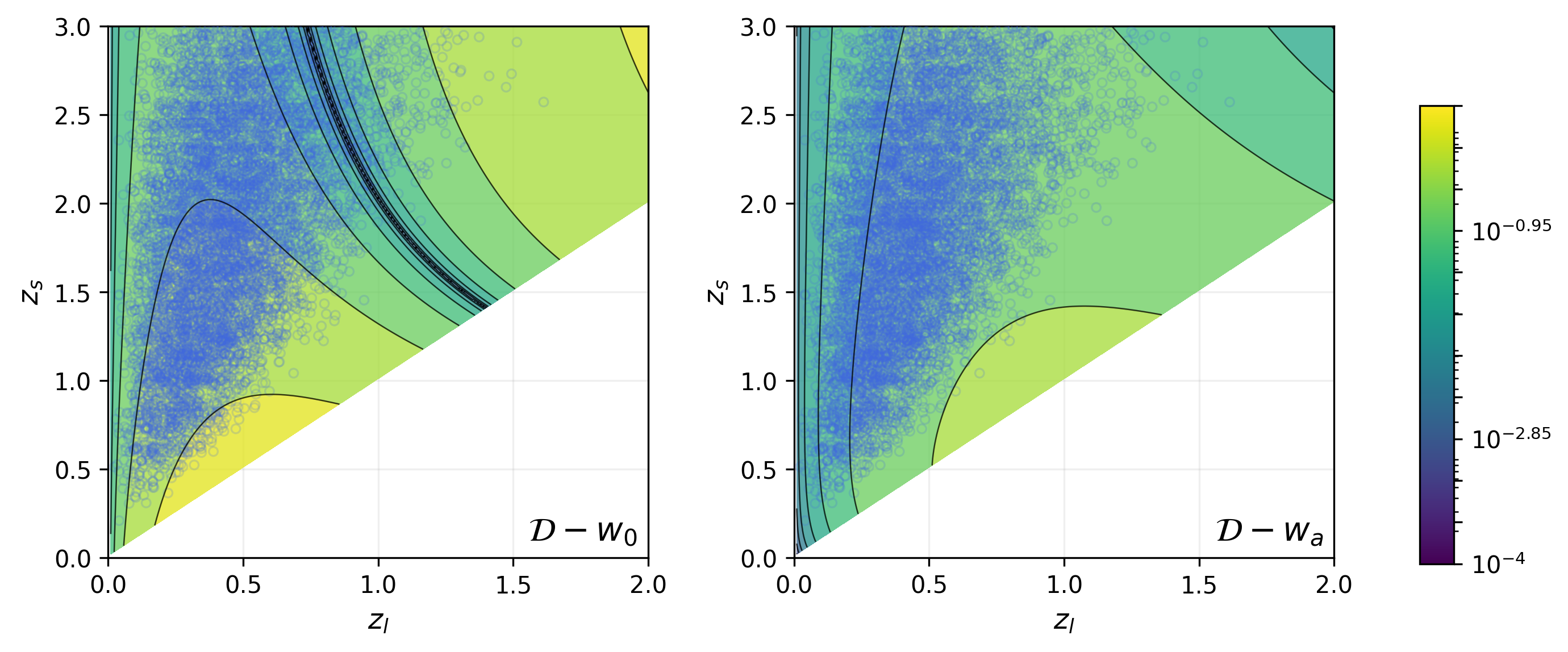}
\caption{\label{fig:fisher R scatter}Sensitivity maps for $w_0$ and $w_a$ assessed using the distance ratio. Blue circles indicate the redshift distribution of the simulation data based on \citet{collett2015population}.}
\end{figure*}

The first row of Fig.~\ref{fig:fisher analysis} illustrates how the sensitivity to cosmological parameters varies with lens and source redshift when the distance ratio is used as the observable. At fixed source redshift, increasing the lens redshift $z_l$ generally enhances the sensitivity to $\Omega_m$, whereas at fixed $z_l$, increasing the source redshift $z_s$ weakens the sensitivity to $\Omega_m$. The parameters $w_0$ and $w_a$ exhibit the similar trends, with higher sensitivity when the lens and source redshifts are closer to each other. For $w_0$, the sensitivity maps display pronounced valleys, indicating a sharp reduction in constraining power. The origin of this valley feature is discussed in detail in Section~\ref{sec:sensitivity valley}. As a result, the sensitivity to $w_0$ displays two local maxima: one at high $z_s$ with high $z_l$, and another around $z_l \sim 0.5$. While the most sensitive regime for $w_a$ is broader, spanning approximately $0.5 \lesssim z_l \lesssim 1.5$. The locations of these low-sensitivity valleys depend on the fiducial cosmological parameters adopted. Increasing $\Omega_m$ or $w_a$ shifts the $w_0$ sensitivity valley toward higher redshifts, whereas increasing $w_0$ shifts it toward lower redshifts. For the time-delay distance, the subplots of $\Omega_m$ and $w_a$ show the low-sensitivity valleys as well, but the overall trend in all three parameters indicates an increasing sensitivity with redshift. As shown in the second row of Fig.~\ref{fig:fisher analysis}, the low-sensitivity valleys for the time-delay distance occur at lens redshifts $z_l \lesssim 0.5$ for $\Omega_m$ and at $0.25 \lesssim z_l \lesssim 1.0$ for $w_a$. Fixing the higher-redshift source at $z_{s_2}=3.0$ and varying the lower-redshift source, we find that the sensitivity in DSP case increases as $z_{s_1}$ approaches $z_l$. Peak sensitivity region for $\Omega_m$ and $w_a$ occurs near $z_l\sim1.0$, whereas for $w_0$ the most sensitive range shifts to lower lens redshift, around $z_l\sim0.75$. As shown in Fig.~\ref{fig:fisher R scatter}, the redshift distribution of our simulated sample based on \citet{collett2015population} substantially overlaps with the high-sensitivity regions for $w_0$ and $w_a$ when the distance ratio is used as the observable.

\begin{figure*}
\centering
\includegraphics[width=0.95\linewidth]{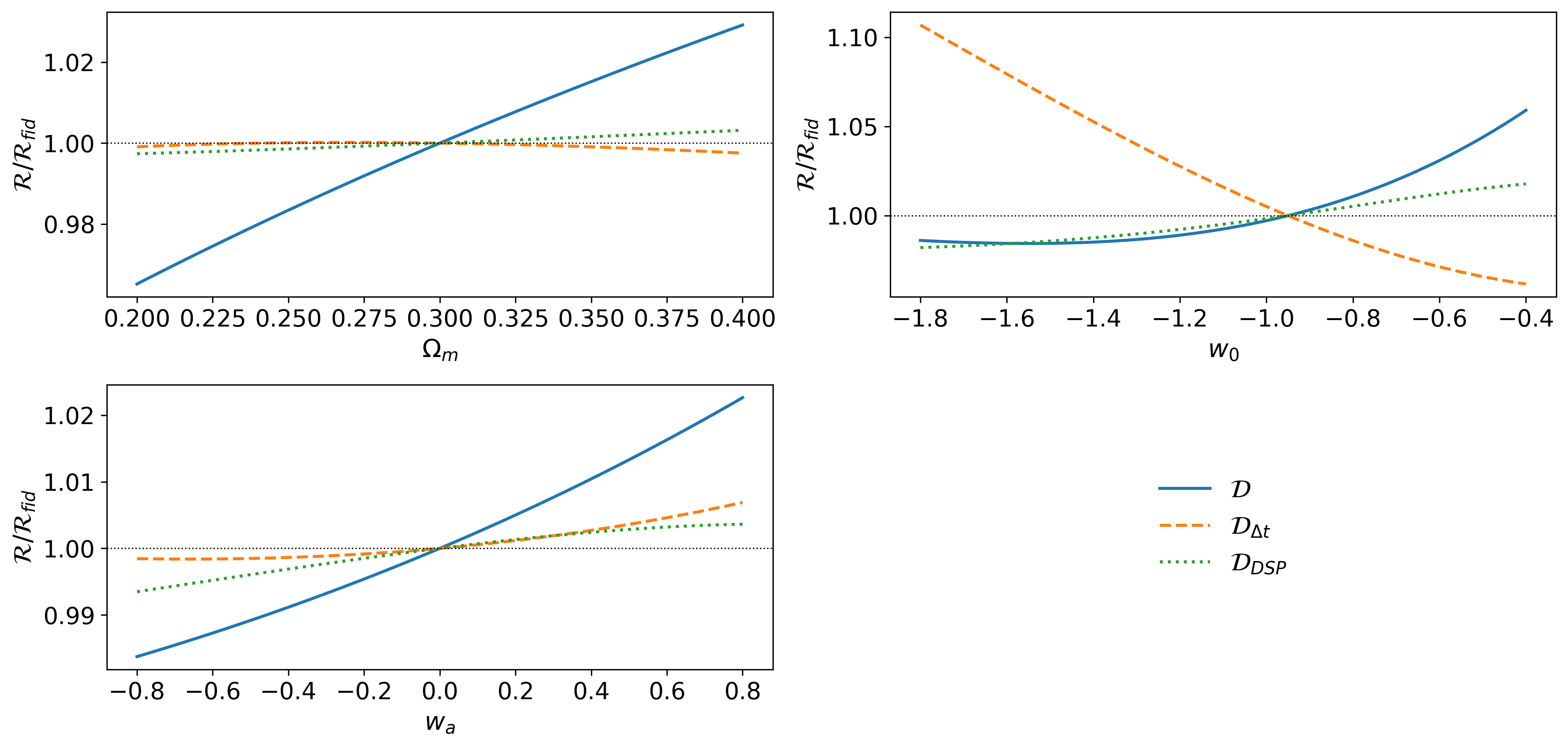}
\caption{\label{fig:R deviation}These three subplots quantify the response of the observables to perturbations of the input cosmological parameters $\Omega_m$, $w_0$, and $w_a$ away from a fiducial model. Specifically, they show how the distance ratio $\mathcal{D}$, the time-delay distance $D_{\Delta t}$, and the double–source–plane distance ratio $\mathcal{D}_{\mathrm{DSP}}$ vary under parameter shifts. Curves are expressed as fractional changes relative to the fiducial value, $\mathcal{R}/\mathcal{R}_{\mathrm{fid}}$, so larger departures potentially indicate higher sensitivity.
}
\end{figure*}

We further examine the degeneracy of each observable from its fiducial value, $\mathcal{R}/\mathcal{R}_{\mathrm{fid}}$, evaluated at $z_l=0.5$, $z_s=1.0$, and for DSP at $z_{s_2}=2.0$, with $\Omega_m=0.3$, $w_0=-0.95$, and $w_a=0.0$. Varying one cosmological parameter away from this fiducial model, the local slope $\partial \left(\mathcal{R}/\mathcal{R}_{\mathrm{fid}}\right)/\partial \lambda_i$ about the fiducial point (with $\lambda_i\in\{\Omega_m,w_0,w_a\}$) serves as a direct proxy for sensitivity. As shown in Fig.~\ref{fig:R deviation} The distance ratio is most sensitive to $\Omega_m$ and $w_a$, whereas the time-delay distance is more sensitive to $w_0$. Although higher sensitivity is able to deliver tighter constraints, it also requires competitive precise measurements of the observables. For example, a $1\%$ uncertainty in the distance ratio for a single system translates into parameter ranges of roughly $0.27 \lesssim \Omega_m \lesssim 0.33$, $-1.2 \lesssim w_0 \lesssim -0.9$, and $-0.5 \lesssim w_a \lesssim 0.4$.

\subsection{Hierarchical Strong Lensing Cosmography}\label{sec:Hierarchical_framework}
This part is devoted exclusively to the technique of using SGL distance ratios as cosmological probes. A central challenge for strong-lensing cosmography is that precise cosmological inference relies on accurate lens-mass modeling, yet many pipelines adopt a fiducial cosmology when constructing the mass model \citep{etherington2022automated,tan2023project}. This practice may introduce circularity and complicates efforts to obtain fully independent constraints. Allowing both the global mass density slope of lens galaxies and cosmological parameters to vary can alleviate this issue, but assuming a single, universal mass profile for all lenses is typically too simplistic. These considerations motivate hierarchical frameworks. Developing such methods will enable more realistic calibration of lens models and yield more robust cosmological constraints from forthcoming large strong-lens samples.

From unanchored SNe\,Ia, we first reconstruct the luminosity distance $D^{L}(z;\omega)$ using a non-parametric regressor (ANN) with weights $\omega$, following \citet{geng2025investigating}.
Via Etherington’s relation $D^L(z)=(1+z)^2 D^A(z)$, the dimensionless distance ratio of lens $i$ at $(z_{{l},i},z_{{s},i})$ is
\begin{equation}
\mathcal D_i(\omega)=\frac{D^A(z_{{l},i},z_{{s},i};\omega)}{D^A(z_{{s},i};\omega)}=1-\frac{1+z_s}{1+z_l}\frac{D^L_{{l},i}}{D^L_{{s},i}},
\end{equation}
in the flat limit $\Omega_k=0$. 
With the SNe\,Ia dataset $\mathcal H=\{(\mu_j,z_j,\sigma_{\mu,j})\}_{j=1}^{N_{\rm SN}}$ and the light-curve standardization nuisance parameters $\psi=\{M_B,\alpha_{\rm SN},\beta_{\rm SN},\sigma_{\rm int}\}$, the SNe--ANN posterior is
\begin{equation}
\begin{aligned}
p(w,\psi\mid \mathcal H)
\propto\ &\prod_{j=1}^{N_{\rm SN}}
\mathcal N \Big(\mu_j;\,5\log_{10} D^L(z_j;w)+M_B,\ \sigma_{\mu,j}\Big)\, \\
p(w)\,p(\psi),
\end{aligned}
\end{equation}
where $\mu_j$ is the distance modulus, $z_j$ is the SN redshift, and $\sigma_{\mu,j}$ is the uncertainty in the distance modulus. The parameter $M_B$ is the SN absolute magnitude, $\sigma_{\rm int}$ is the intrinsic scatter, and $\alpha_{\rm SN}$ and $\beta_{\rm SN}$ are the light-curve shape and color correction coefficients, respectively. The induced cosmology-agnostic posterior for the distance ratio of lens $i$ is
\begin{equation}
p(\mathcal D_i\mid \mathcal H)=\iint \delta\!\left[\mathcal D_i-\mathcal D_i(w)\right]\,
p(w,\psi\mid \mathcal H)\,dw\,d\psi,
\label{eq:R_from_SNe}
\end{equation}
which is independent of the absolute calibration, because $M_B$ (or equivalently the $H_0$ degeneracy) cancels in $\mathcal D_i$. This cosmology-agnostic posterior is then used as a prior to calibrate the lens-population parameters. For lens $i$, we collect the individual parameters $\xi_i = \{\gamma_i,\delta_i,\beta_i\}$ and the population hyper–parameters $\eta = \{\gamma_0,\gamma_s,\sigma_{\gamma}, \delta_0,\delta_s,\sigma_{\delta}, \beta_0,\sigma_{\beta}\}$. The total mass–density slope follows $\gamma_i \sim \mathcal{N} \big(\gamma_0 + \gamma_s \times(z_i-z_{\mathrm{med}},\ \sigma_{\gamma}^2\big)$, and the luminous mass density slope follows $\delta_i \sim \mathcal{N} \big(\delta_0 + \delta_s \times(z_i-z_{\mathrm{med}},\ \sigma_{\delta}^2\big)$. The stellar velocity–dispersion anisotropy parameter obeys $\beta_i \sim \mathrm{Tri}\big(-0.5,\,0.656;\ \mathrm{mode}=0.102\big)$ for the triangular prior from \citet{geng2025investigating}.

Introducing cosmological parameters $\lambda=(\Omega_m,w)$ or
$\lambda=(\Omega_m,w_0,w_a)$, 
the theoretical angular–diameter distance ratio for lens $i$ is
\begin{equation}
\mathcal D_i^{\rm th}(\lambda)=
\frac{D^A(z_{{l},i},z_{{s},i};\lambda)}
     {D^A(z_{{s},i};\lambda)}.
\label{eq:Dth_def}
\end{equation}
In the fully hierarchical analysis we infer $\lambda$ and $\eta$
simultaneously from the supernova and strong–lensing data by sampling
the joint posterior
\begin{equation}
\begin{aligned}
&p\left(\lambda,\eta,\{\xi_i\},\{\mathcal D_i\}\mid
          \{\theta_{E,i},\sigma_{ap,i}\},\mathcal H\right)\\
&\propto
p(\lambda)\,p(\eta)\,
\prod_{i=1}^{N}
\Big[
  p \left(\theta_{E,i},\sigma_{ap,i}\mid \xi_i,\mathcal D_i\right)\,
  p(\xi_i\mid\eta)\,
  p \left(\mathcal D_i\mid\mathcal H\right)\, \\
  &\quad\quad\quad \times \delta \big(\mathcal D_i-\mathcal D_i^{\rm th}(\lambda)\big)
\Big],
\end{aligned}
\label{eq:full_hier_joint}
\end{equation}
where $p(\mathcal D_i\mid\mathcal H)$ is the SNe–ANN distance–ratio
posterior given in Eq.~\eqref{eq:R_from_SNe}, and the Dirac delta
enforces consistency between the latent distance ratio $\mathcal D_i$
and the theoretical prediction $\mathcal D_i^{\rm th}(\lambda)$.

The cosmological posterior is obtained by marginalizing over all
lens–level and population–level quantities,
\begin{equation}
\begin{aligned}
p& \left(\lambda\mid
         \{\theta_{E,i},\sigma_{ap,i}\},\mathcal H\right) \\
&\propto
p(\lambda)\int d\eta
\prod_{i=1}^{N}
\iint d\mathcal D_i\,d\xi_i\;
p \left(\theta_{E,i},\sigma_{ap,i}\mid \xi_i,\mathcal D_i\right)\\
&\qquad\times
p(\xi_i\mid\eta)\,
p \left(\mathcal D_i\mid\mathcal H\right)\,
\delta \big(\mathcal D_i-\mathcal D_i^{\rm th}(\lambda)\big)\,
p(\eta).
\end{aligned}
\label{eq:full_hier_cosmo_post}
\end{equation}
Eqs.~\eqref{eq:full_hier_joint}--\eqref{eq:full_hier_cosmo_post}
constitute a fully hierarchical formulation. The cosmological parameters $\lambda$, the population hyper–parameters $\eta$, and the lens–level parameters $\{\xi_i\}$ are inferred in a single Bayesian framework, while the SNe–ANN uncertainties on the distance ratios enter through the prior $p(\mathcal D_i\mid\mathcal H)$ rather than via a separate calibration stage.

In our framework, Type~Ia supernovae are not used to directly impose a parametric dark-energy model. Instead, we first reconstruct cosmology-agnostic posteriors for the lensing distance ratios $\mathcal{D}_i$ at the lens redshifts. We then combine these distance-ratio posteriors with strong-lensing observables to build priors only on the redshift evolution and intrinsic scatter of the lens mass-density slopes for the cosmological stage. In the cosmological model-testing stage, we jointly sample the lens-population parameters ($\eta$) and the cosmological parameters ($\lambda$).

\section{Results and discussion} \label{sec:results}

In this section, we present the constraints from the strong-lensing distance ratios alone, as well as the joint constraints obtained by combining them with \textit{Planck}. The results are summarized in Table~\ref{tab:cosmo-results}.

\begin{deluxetable*}{lllll}
\tablecaption{Constraints on cosmological model parameters obtained with SGL distance ratio alone and jointly with CMB data and with CMB+BAO data. 
\label{tab:cosmo-results}
}
\tablewidth{0pt}
\tablehead{
\colhead{Model} & \colhead{Datasets} & \colhead{$\Omega_m$} & \colhead{$w$ or $w_0$} & \colhead{$w_a$}
}
\startdata
Flat $w\mathrm{CDM}$ & Strong Lensing & $0.32_{-0.11}^{+0.10}$ & $-1.00_{-0.97}^{+0.57}$ & -- \\
Flat $w\mathrm{CDM}$ & Strong Lensing + \textit{Planck} & $0.256_{-0.039}^{+0.042}$ & $-1.142_{-0.083}^{+0.080}$ & -- \\
Flat $w_0w_a\mathrm{CDM}$ & Strong Lensing & $0.348_{-0.099}^{+0.099}$ & $-1.22_{-1.21}^{+0.79}$ & $-1.5_{-3.1}^{+3.4}$ \\
Flat $w_0w_a\mathrm{CDM}$ & Strong Lensing + \textit{Planck} + BAO & $0.331_{-0.023}^{+0.024}$ & $-0.68_{-0.25}^{+0.26}$ & $-1.03_{-0.78}^{+0.70}$ \\
\enddata
\end{deluxetable*}

\begin{figure}
\centering
\includegraphics[width=0.9\linewidth]{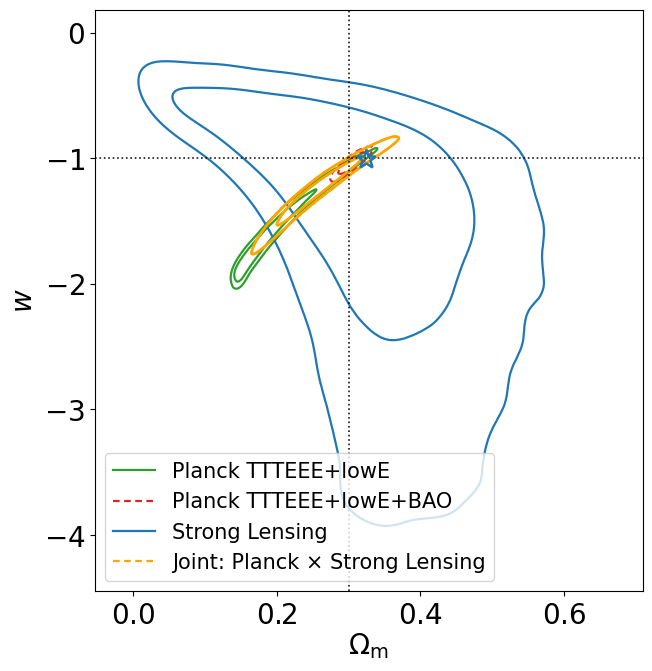}
\caption{\label{fig:flat wCDM} Joint constraints on $\Omega_m$ and $w$.
Blue solid contours show constraints from strong-lensing distance ratios alone. 
Green dashed contours show \textit{Planck}-only constraints.  
Orange solid contours show the joint constraints from strong-lensing distance ratios + \textit{Planck}.
}
\end{figure}

\begin{figure}
\centering
\includegraphics[width=0.9\linewidth]{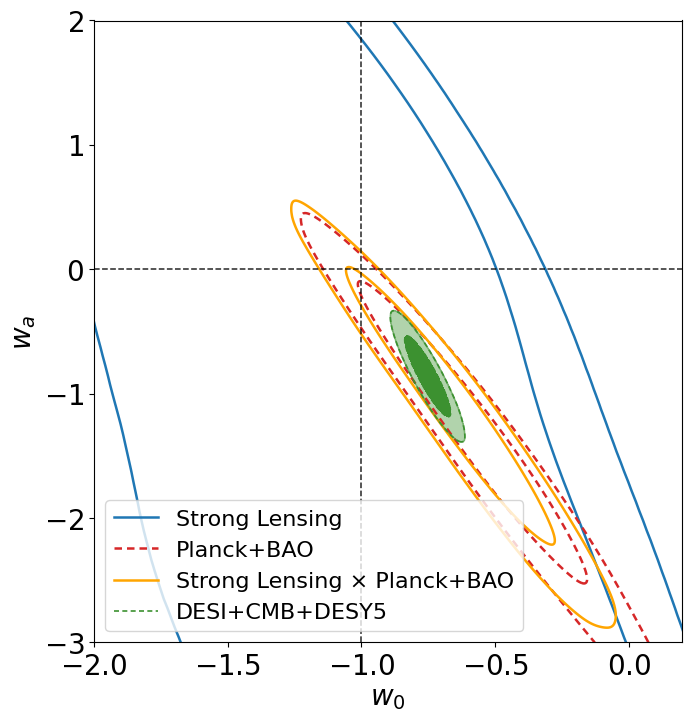}
\caption{\label{fig:flat w0waCDM} Joint constraints on $w_0$ and $w_a$.
Blue solid contours show constraints from strong lensing distance ratios alone. 
Red dashed contours show \textit{Planck}+BAO constraints.  
Orange solid contours show the joint constraints from strong-lensing distance ratios + \textit{Planck} + BAO. Green filled contours show the joint constraints from DESI+CMB+DES Year~5 \citep{desicollaboration2025desi}.
}
\end{figure}

Cosmological constraints from current data: for the flat $w$CDM model, the marginalized constraints in the $w_0$--$w_a$ plane are shown in Fig.~\ref{fig:flat wCDM}, and the full joint posteriors are shown in Fig.~\ref{fig:flat wCDM vs w0wa}. Accounting for the redshift evolution of the lens-galaxy mass–density slope in the combined strong-lensing (SGL) sample, we obtain $\Omega_m = 0.32_{-0.11}^{+0.10}$ and $w = -1.00_{-0.97}^{+0.57}$. When combined with \textit{Planck}, the constraints tighten significantly due to the complementary degeneracy directions:
$\Omega_m = 0.256_{-0.039}^{+0.042}$ and $-1.142_{-0.083}^{+0.080}$. Compared with SGL alone, the joint analysis improves the precision on both $\Omega_m$ and $w$ by a factor of $\sim 2$–$3$, illustrating the validity of combining geometric information from SGL with the CMB. In the flat $w_0 w_a$CDM model we adopt flat priors for the dark-energy equation-of-state parameters, taking $\mathcal{U}(-3,1)$ for $w_0$ and $\mathcal{U}(-5,5)$ for $w_a$. Accounting for the redshift evolution of the lens-galaxy mass–density slope in the combined strong-lensing sample, we obtain
$w_0 = -1.22_{-1.21}^{+0.79}$ and $w_a = -1.5_{-3.1}^{+3.4}$. The SGL-only posteriors remain broad, which reflects the well-known degeneracy between $w_0$ and $w_a$ in geometric probes as well as residual correlations with lens-structure parameters. When we combine the SGL data with \textit{Planck} results, the constraints tighten to $w_0 = -0.68_{-0.25}^{+0.26}$ and $w_a=-1.03_{-0.78}^{+0.70}$. The joint result is consistent with, and closely reproduces, the DESI+CMB+DES Year~5 constraints \citep{desicollaboration2025desi}, which is shown by green contours in Fig.~\ref{fig:flat w0waCDM}.

Cosmological constraints with simulated data: For the simulated sample, we constrain the parameter sets $(\Omega_m, w)$ and $(\Omega_m, w_0, w_a)$. The fiducial values and our results are listed in Table~\ref{tab:cosmo-input}. The joint posteriors of the cosmological parameters and the lens-population parameters are shown in Fig.~\ref{fig:flat wCDM sim} for the $w$CDM model and in Fig.~\ref{fig:flat w0waCDM sim} for the $w_0w_a$CDM model. With 7500 strong lenses and their kinematic measurements, we obtain tight constraints: $\Delta \Omega_m \simeq 0.01$ and $\Delta w \simeq 0.1$. The posteriors recover the fiducial input values of $\Omega_m$ and $w$. We also find a strong degeneracy between $\gamma_0$ and $\delta_0$. For the $w_0w_a$CDM model, the fiducial input values of $\Omega_m$, $w_0$, and $w_a$ are also well recovered. Strong degeneracies among the cosmological parameters are visible. Overall, in both the $w$CDM and $w_0w_a$CDM models, the cosmological parameters are recovered close to their fiducial values, while the lens-population parameters show small biases.

Now we will discuss in more details several aspects of our methods and results. Section~\ref{sec:sensitivity valley} explains the origin of the valley feature in the sensitivity to cosmological parameters. Section~\ref{sec: density evo} discusses why the redshift evolution of the lens-galaxy density slope should be included in cosmological analyses. Section~\ref{sec: perspective of dis ratio} provides a broader perspective on distance-ratio cosmography.

\begin{deluxetable}{lcccc}
\tablecaption{Constraints of cosmological parameters and lens population parameter under simulation data.\label{tab:cosmo-input}}
\setlength{\tabcolsep}{4pt} 
\tablewidth{0pt} 
\tablehead{
\colhead{Parameter} & \colhead{Fiducial($w$CDM)} & \colhead{Fiducial($w_0w_a$CDM)} &
\colhead{Flat $w\mathrm{CDM}$} & \colhead{Flat $w_0w_a\mathrm{CDM}$}
}
\startdata
$\Omega_m$                & 0.3                          & 0.3                            & $0.293^{+0.012}_{-0.0089}$     & $0.285^{+0.075}_{-0.040}$      \\
$w$ or $w_0$              & -1                           & -0.75                          & $-1.04^{+0.12}_{-0.11}$        & $-0.70^{+0.19}_{-0.41}$         \\
$w_a$                     & --                           & -0.86                          & --                             & $-1.14^{+2.2}_{-0.91}$         \\
\tableline 
$\gamma_0$                & 2.054                        & 2.054                          & $2.0500^{+0.0027}_{-0.0027}$   & $2.0499^{+0.0028}_{-0.0028}$   \\
$\gamma_s$                & -0.18                        & -0.18                          & $-0.2015^{+0.0064}_{-0.0064}$  & $-0.2015^{+0.0064}_{-0.0064}$  \\
$\delta_0$                & 2.26                         & 2.26                           & $2.2389^{+0.0093}_{-0.011}$    & $2.2387^{+0.0094}_{-0.011}$    \\
$\delta_s$                & -0.16                        & -0.16                          & $-0.153^{+0.011}_{-0.011}$     & $-0.152^{+0.011}_{-0.011}$     \\
$\beta_0$                 & 0.102                        & 0.102                          & $0.10199^{+0.00014}_{-0.00014}$& $0.10199^{+0.00014}_{-0.00014}$\\
\enddata
\tablecomments{Table shows the fiducial values and constrained results of cosmological parameters ($\Omega_m, w/w_0, w_a$) and lens population parameters ($\gamma_0, \gamma_s, \delta_0, \delta_s, \beta_0$) from simulation data.}
\end{deluxetable}

\subsection{Sensitivity valley}\label{sec:sensitivity valley}
In the section~\ref{sec:sensitivity analysis}, we found that for some parameter space there would be a low-sensitivity valley, where the constraining sensitivity drops to 0. This valley stems from the non-monotonic feature of the derivative of angular-diameter distance function. The key part of sensitivity factor, which is the partial derivative part, for distance ratio can be written as
\begin{equation}
\begin{aligned}
\frac{\partial \ln (D^A_{ls}/D^A_s)}{\partial \lambda_i}
&=\frac{d_l}{d_s - d_l} \left( \frac{\partial_{\lambda_i} d_s}{d_s} - \frac{\partial_{\lambda_i} d_l}{d_l} \right) \\
&=\frac{d_l}{d_s - d_l} \left( g(z_s)-g(z_l) \right),
\end{aligned}
\end{equation}
where $g(z)={\partial_{\lambda_i} d(z)}/{d(z)}$ describes how the comoving distance $d(z)$ at certain redshift change with different cosmological parameters $\lambda_i$. Lowercase notations denote dimensionless distances, with the factor $c/H_0$ scaled out. The function $g(z)$ becomes non-monotonic when $\lambda_i=w_0$ for the distance ratio and when $\lambda_i=\Omega_m$ or $w_a$ for the time-delay distance. In these cases, the sensitivities of $d_{ls}$ and $d_s$ can partially cancel once $z_s$ passes the turning point of $g(z)$. When $g(z_s)$ equals $g(z_l)$, the cancellation is complete. 

The existence of these valleys is consistent with earlier sensitivity analyses of the distance ratio \citep{piorkowska2013complementarity} and the time-delay distance \citep{coe2009cosmological,linder2011lensing,jee2016timedelay}. In particular, \citet{piorkowska2013complementarity} found vanishing sensitivity to $w_0$ in the distance ratio at $z_l \sim 1$ when assuming $z_s = 2 z_l$. For time-delay distance studies, \citet{linder2011lensing} also adopted $z_s = 2 z_l$. The zero-sensitivity locations for $\Omega_m$ and $w_a$ reported by \citet{linder2011lensing} can be recovered in Fig.~\ref{fig:fisher analysis} by taking the intersections between the sensitivity valleys and the line $z_s = 2 z_l$.

The existence of a low-sensitivity valley implies that, for certain $(z_l, z_s)$ combinations, cosmological constraints can remain weak even with arbitrarily precise measurements. As a result, most of the constraining power comes from systems outside the low-sensitivity region. This analysis identifies which redshift ranges dominate cosmological inference when combining galaxy-scale strong-lensing data with different distance combinations in the era of large surveys.

\subsection{Density slope evolution in strong lensing cosmology}\label{sec: density evo}

\begin{figure*}
\centering
\includegraphics[width=0.55\linewidth]{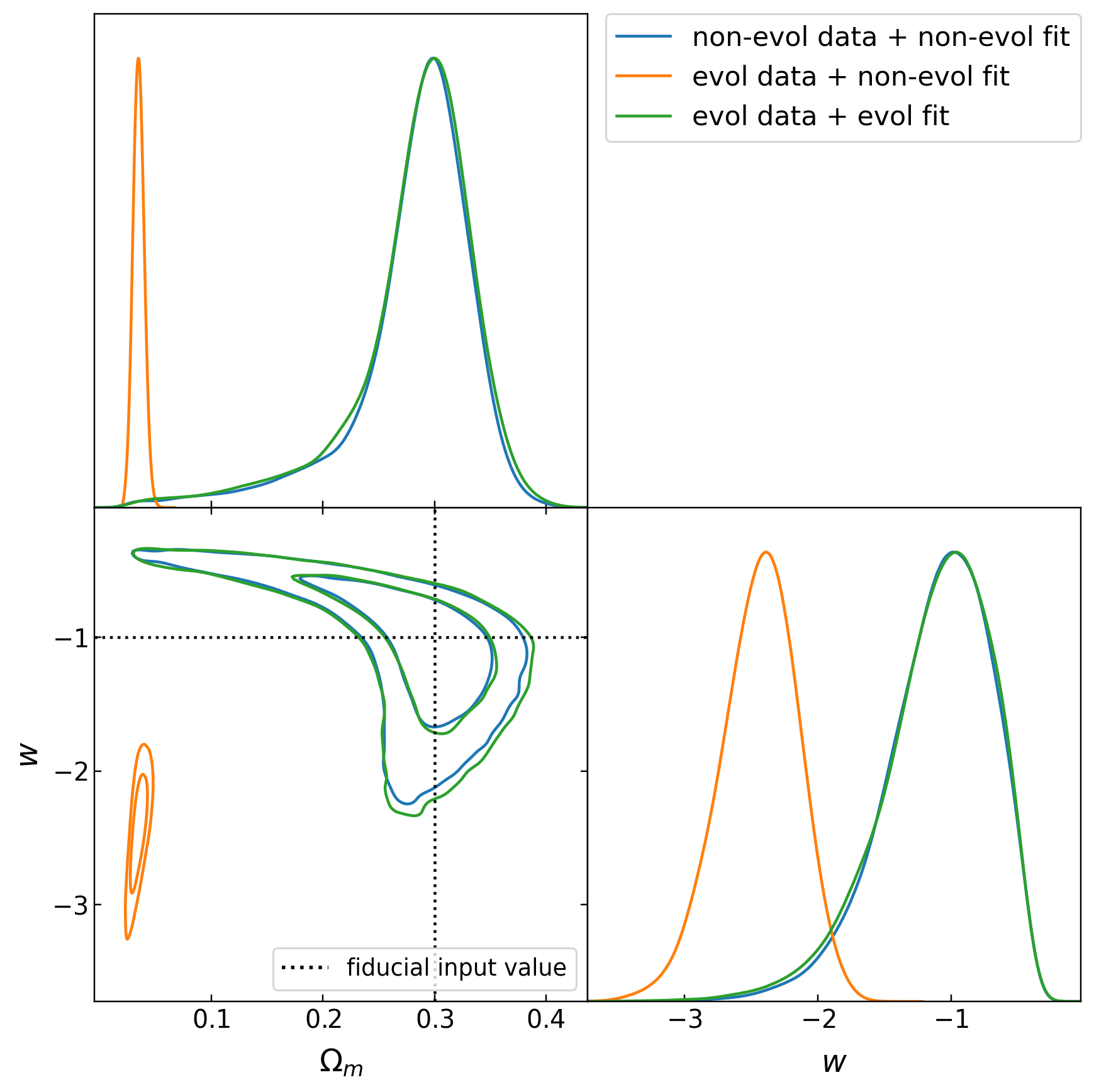}
\caption{\label{fig:sim wCDM} Posterior distribution of the cosmological parameter with fixed lens parameters. Blue contours indicate the cosmological constraints for the  simulation data with non-evolving lens model (single $\gamma$). Green contours indicate the cosmological constraints for the data with evolving lensing model ($\gamma_0$ and $\gamma_s$). The orange contours indicate the cosmological constraints for the evolving lens mass density power-law exponent and simulation data with non-evolving lens model.
}
\end{figure*}

Although the sign of the redshift evolution in the lens-galaxy density slopes is still uncertain \citep{etherington2022bulgehalo,sahu2024agel,geng2025investigating}, one point is clear: if such evolution is present, adopting a single global slope for all lenses will bias cosmological inference. We test this effect using 1000 mock strong-lensing systems with redshift-evolving density slopes generated with the code of \citet{collett2015population}. These systems are fitted with a non-evolving model, with the slope fixed to $\gamma_{\mathrm{tot}}=2.078$ from \citet{auger2010sloanb}. The evolving model is the same as that used for the simulated sample in Section~\ref{sec:simulation data}. In this case, the inferred contours (orange in Fig.~\ref{fig:sim wCDM}) shift by $\sim 10\sigma$ from the true value of $\Omega_m$, even though the 68\% confidence region appears very tight. In contrast, mock data without intrinsic evolution (blue) deliver similar constraining power to the evolving case when the analysis model matches the data.

To account for the mass-density-slope evolution of lens galaxies, we adopt the hierarchical inference framework described in Section~\ref{sec:Hierarchical_framework}. Our results show that this framework can removes the bias induced by slope evolution, while the lens-population parameters still show internal degeneracies, especially between $\gamma_0$ and $\delta_0$. The small residual biases in population parameters, such as $\gamma_0$ and $\delta_0$, are likely caused by these degeneracies. In our framework, the SN~Ia information is used only to set priors on the lens-population parameters and does not directly constrain the cosmological parameters. This is sufficient to recover the population parameters at an approximate level and to recover the cosmological parameters accurately. However, breaking the degeneracy between the total-mass and luminous-mass components will likely require additional observations.

\subsection{Challenge and perspective of distance ratio cosmology}\label{sec: perspective of dis ratio}

Compared with studies that assume a single global total mass--density slope \citep{qi2022cosmological}, our hierarchical framework improves the precision of cosmological constraints by about one order of magnitude. It also reaches a better precision to works that either fix the luminous-matter density slope or fit fewer cosmological parameters \citep{chen2019assessing,li2024cosmology}. This improvement highlights the need to break the degeneracy between cosmological parameters and mass-profile evolution. We also confirm that strong lensing and \textit{Planck 2018} data show nearly orthogonal degeneracy directions in the $\Omega_m$--$w$ plane under $w$CDM. These results demonstrate the complementarity between SGL and CMB information, and the value of population-level lens modeling when inferring the evolution of the dark-energy equation of state.

The theoretical analysis in Section~\ref{sec:sensitivity analysis} shows that distance-ratio cosmography is more sensitive to most parameters in models beyond $\Lambda$CDM. In practice, however, hundreds of distance-ratio lenses give a precision comparable to that from only eight time-delay systems in \citet{tdcosmocollaboration2025tdcosmo}. A key reason is the measurement precision. Time delays can reach $\sim 1\%$ in favorable cases, while distance-ratio inferences have larger observational errors. The dominant contribution comes from the stellar velocity dispersion, whose uncertainty propagates directly into the cosmological constraints.

There are two main ways to reduce this limitation. First, more informative stellar-kinematic data (e.g., integral-field-unit spectroscopy - IFU) can better constrain the dynamical modeling of lens galaxies. Second, a larger sample can average down measurement noise. The second option is more feasible for current and upcoming surveys, since IFU spectroscopy is hard to scale to very large samples, while LSST and related programs will likely discover a significant number of galaxy-scale lenses. In addition, because SN~Ia and strong-lensing distance ratios have different degeneracy directions in cosmological parameter space \citep{biesiada2011dark,piorkowska2013complementarity,cao2015cosmology}, the growing SN~Ia samples will further help break the degeneracy between cosmological parameters and mass-profile evolution, leading to tighter cosmological constraints.

Line–of–sight contamination and selection effects are additional challenges for distance–ratio cosmography. The internal mass–sheet degeneracy can be reduced using velocity dispersion information, but the external convergence $\kappa_{\rm ext}$, driven by unmodeled line–of–sight mass, remains difficult to control. In this work we do not apply an explicit $\kappa_{\rm ext}$ correction. This omission can introduce a systematic in the inferred redshift evolution and, in turn, in the final cosmological constraints. Future analyses should include environment reconstructions and selection–function modeling to address these systematics.

\section{Summary}\label{sec:summary}

In this paper, we present a Fisher-like 
sensitivity analysis of cosmological parameters. For the distance ratio, we find low-sensitivity valleys in $w_0$ and $\alpha$. For the time-delay distance, valleys appear in $\Omega_m$, $w_a$, and $\alpha$. At these valleys, the derivative of the observable with respect to the parameter vanishes, so the sensitivity to that parameter drops to zero. This happens regardless of the measurement precision or the details of lens modeling. 

Motivated by our previous work \citep{geng2025investigating}, we develop a hierarchical inference framework to calibrate the redshift evolution of mass-density slopes and to constrain cosmology. We use unanchored Type~Ia supernova data and a non-parametric artificial neural network to derive priors on the mass-profile evolution of lens galaxies. Using strong-lensing observables, we calculate distance ratios and compare them with theoretical predictions for a given cosmology. This framework infers the redshift evolution of the lens mass profile and constrains the cosmological model at the same time. We test the method with mock data based on LSST forecasts. These tests show the expected gain in constraining power once the mass-profile evolution is calibrated. Based on our results, we draw the main conclusions as follows.

\begin{enumerate}

\item The distance ratio is sensitive to $\Omega_m$, $w$ (or $w_0$), and $w_a$. Most strong-lensing systems expected from LSST lie outside the distance-ratio low-sensitivity valleys and therefore retain strong constraining power. Large-field surveys will produce large samples of galaxy--galaxy lenses, which strengthens the promise of distance-based cosmology.\\

\item The hierarchical framework constrains the redshift evolution of the lens mass profile with minimal assumptions, while extracting cosmological information from galaxy--galaxy lenses. Adding external data further breaks parameter degeneracies across probes. Joint constraints from strong-lensing distance ratios and \textit{Planck} CMB data yield $\Omega_m=0.256^{+0.042}_{-0.039}$ and $w=-1.142^{+0.080}_{-0.083}$ under flat $w$CDM. Constraints combining \textit{Planck} and BAO give $\Omega_m=0.331^{+0.024}_{-0.023}$ and $w_0=-0.68^{+0.26}_{-0.25}$ under flat $w_0w_a$CDM, consistent with DESI+CMB+DES Year~5 \citep{desicollaboration2025desi}. For $\sim 10\,000$ simulated lenses, the $w$CDM constraints reach $\Delta\Omega_m \simeq 0.01$ and $\Delta w \simeq 0.1$.\\

\item Characterizing redshift evolution in strong-lensing systems is critical for the distance-ratio cosmography. In simulations with velocity-dispersion measurements and evolving mass profiles, enforcing a non-evolving, uniform profile can bias $\Omega_m$ by up to $10\sigma$. Using the correct redshift-evolution prior recovers the fiducial input values.\\
\end{enumerate}

Summarizing, theoretical, observational, and simulation studies show that strong-lensing distance ratios in the role of cosmological probes have strong intrinsic leverage, due to favorable degeneracy directions and, importantly, good scalability with the sample size. As large galaxy surveys will deliver much larger lens samples, population-level effects, including redshift evolution in the lens-galaxy mass-density slope, will become increasingly important for accurate and precise cosmological inference. In this context, a hierarchical Bayesian framework that jointly models lens-population parameters and cosmological parameters will be essential for using strong lensing to explore the expansion history of the Universe.

\begin{acknowledgements}
      We would like to thank Yiping Shu and Phil Marshall for their helpful suggestions. This research was supported by Beijing Natural Science Foundation No. 1242021; the National Natural Science Foundation of China (Nos. 12021003, 12203009, 12433001); the program of China Scholarships Council. M.B. was supported by the Polish National Science Centre grant 2023/50/A/ST9/00579. M.B. gratefully acknowledges the support from COST Action CA21136 – “Addressing observational tensions in cosmology with systematics and fundamental physics (CosmoVerse)”.
\end{acknowledgements}


\software{astropy \citep{astropycollaboration2013astropy,astropycollaboration2018astropy,astropycollaboration2022astropy},  
          jax \citep{bradbury2021jax},
         numpyro \citep{phan2025numpyro}
          }

\clearpage
\appendix
The appendix presents posterior distribution plots for the lens-population and cosmological parameters, based on the observed data and the simulations, respectively.

\begin{figure*}
\centering
\includegraphics[width=0.9\linewidth]{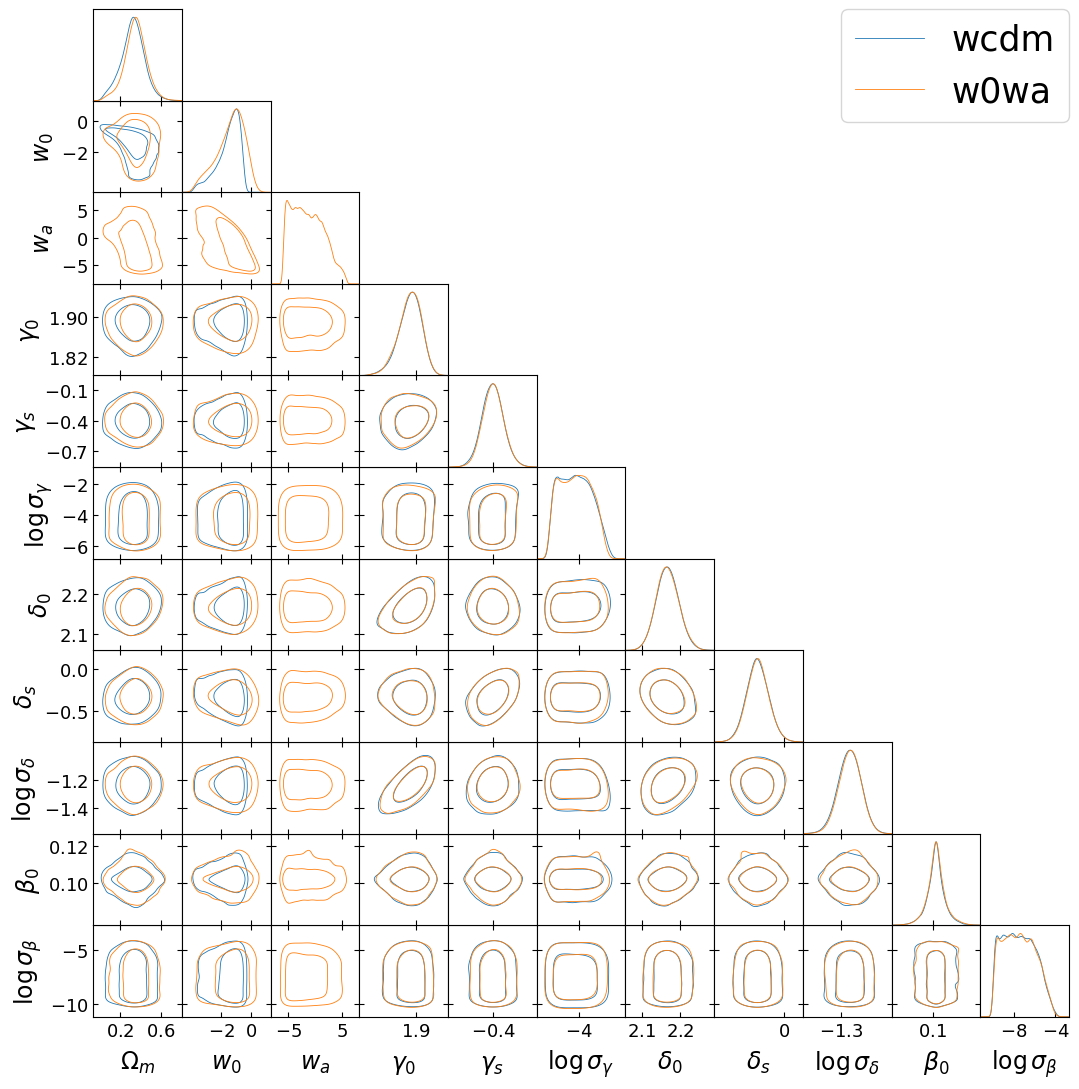}
\caption{\label{fig:flat wCDM vs w0wa}Posterior distributions of the cosmological and lens-population parameters under the $w$CDM and $w_0w_a$CDM models, based on the observed lens-galaxy sample of \citet{chen2019assessing}.
}
\end{figure*}

\begin{figure*}[!t]
\centering
\includegraphics[width=0.9\linewidth]{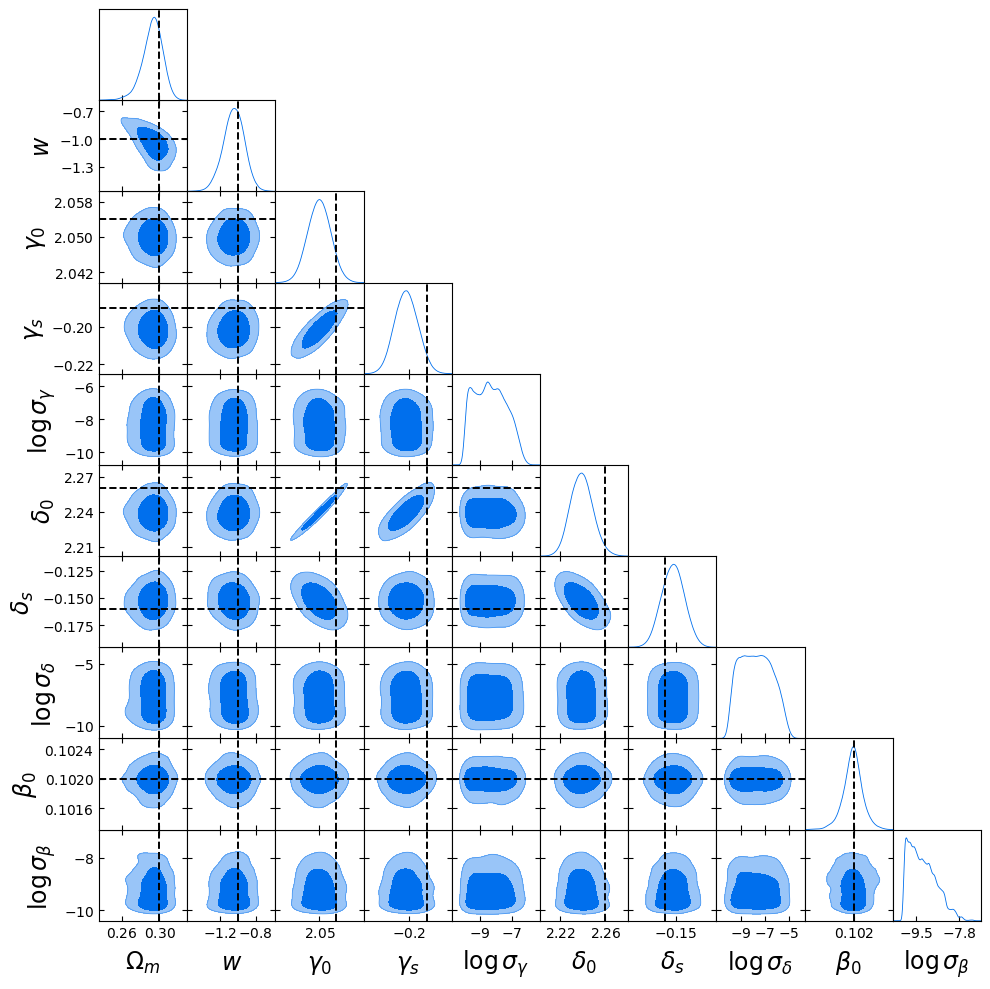}
\caption{\label{fig:flat wCDM sim}Posterior distributions of the cosmological and lens-population parameters under the $w$CDM model, based on 7500 simulated lensing systems.}
\end{figure*}

\begin{figure*}[!t]
\centering
\includegraphics[width=0.9\linewidth]{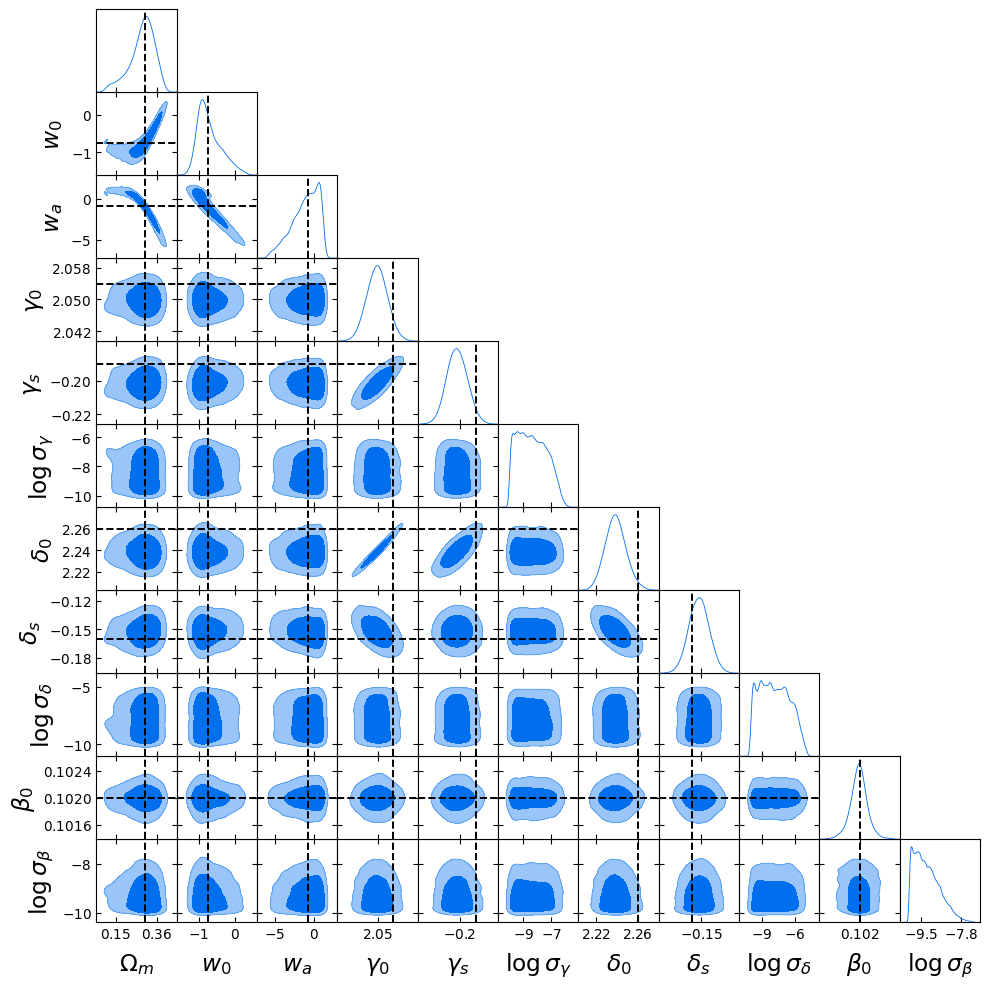}
\caption{\label{fig:flat w0waCDM sim}Posterior distributions of the cosmological and lens-population parameters under the $w_0w_a$CDM model, based on 7500 simulated lensing systems.}
\end{figure*}

\clearpage

\bibliography{ref_gamma}{}
\bibliographystyle{aasjournalv7}



\end{document}